# High performance position-sensitive-detector based on graphene-silicon heterojunction

Wenhui Wang[1], Zhenzhong Yan[1], Jinfeng Zhang[2], Junpeng Lu[1], Hua Qin[2], and Zhenhua Ni[1,*]

[1] School of Physics and Key Laboratory of MEMS of the Ministry of Education, Southeast University, Nanjing 211189, China.

[2] Key Laboratory of Nanodevices and Applications, Suzhou Institute of Nano-tech and Nano-bionics (SINANO), Chinese Academy of Sciences, 398 Ruoshui Road, Suzhou, 215123, China

**Position-sensitive-detectors (PSDs) based on lateral photoeffect have been widely used in diverse applications, including optical engineering, aerospace and military fields. With increasing demands in long working distance, low energy consumption, and weak signal sensing systems, the poor responsivity of conventional Silicon-based PSDs has become a bottleneck limiting their applications. Herein, we propose a high-performance passive PSD based on graphene-Si heterostructure. The graphene is adapted as a photon absorbing and charge separation layer working together with Si as a junction, while the high mobility provides promising ultra-long carrier diffusion length and facilitates large active area of the device. A PSD with working area of 8 mm × 8 mm is demonstrated to present excellent position sensitivity to weak light at nWs level (much better than the limit of ~µWs of Si p-i-n PSDs). More importantly, it shows very fast response and low degree of non-linearity of ~3%, and extends the operating wavelength to the near infrared (IR) region (1319 and 1550 nm). This work therefore provides a new strategy for high performance and broadband PSDs.**

**1. Introduction**

The precise optical measurement of position, distance, displacement, angle and other relevant physical variables are commonly achieved by position sensitive detector (PSD) using the lateral photoeffect.[1-9] Silicon (Si) p-n or p-i-n junctions[2-6] are the most commonly used structures for current PSD. Photoexcited electron-hole (e-h) pairs are separated by built-in electric field at the junction, and the carriers diffuse in the surface layer and are collected by two terminal electrodes. The current/voltage difference detected by the two electrodes is then used to determine the position of light. The key factors that affect the performance of PSD are the efficiency of photon absorption and carrier separation, as well as the diffusion length of carriers, which are usually improved by applying an electrical bias to the Si p-i-n junction.[3] Nevertheless, the minimum detection power (>μW) of Si based PSD[3] is still depressing and has become the bottleneck limiting its applications. For instance, in laser guiding systems, the effective range of the seeker directly relies on the minimum detection power of the PSD components. PSDs with various architectures have been reported in past decades, such as Ti/Si amorphous superlattices,[7] metal-Si or metal-SiO$_2$-Si,[8-10] and GaAs/AlGaAs junction.[11] These devices are appealing for good linearity[8,9] and fast response, but they possess relatively lower photo responsivity and are not suitable for weak signal detection. Moreover, the operating wavelength of the above mentioned PSDs are generally in visible region,[6] while PSDs that can work at infrared (IR) wavelength are normally based on thermopile detector[12] and has a drawback of slow response speed and requires cooling system.

In this work, we present a broadband and high performance PSD based on the graphene-Si heterojunction. The high mobility[13,14] graphene behaves as a photon absorbing and charge separation layer working together with Si as a junction.[15] More importantly, it also serves as a carrier extraction and transportation layer to ensure the ultra-long diffusion distance of the carriers and the large working area of the device. The graphene based PSD is a passive device, has the power detection limit as low as ~17 nW, non-linearity of ~3% at power of ~10 µW, and large working area of >8 mm × 8 mm. It also extends the operating wavelength of Si based PSDs to the near IR region (up to 1550 nm).

## 2. Principles and Characteristics

### 2.1. Structure and principle of the PSD

**Figure 1**a shows the schematic diagram of the graphene-Si hybrid PSDs. It is constructed by placing a large-area chemical vapor deposition (CVD) grown monolayer graphene onto a lightly n-doped Si ($\rho$ = 1-10 $\Omega$ cm) substrate. Two pairs of Ni (5 nm)/Au (50 nm) electrodes are deposited on graphene for signals collection. The Raman spectrum of graphene is shown in Figure S1 (Supporting information), which suggests the monolayer thickness and high quality of the graphene sample.[16] As shown in Figure 1b, the pinning effect[17] due to the surface states result in the band bending of Si surface layer (depletion layer) and a built-in electric field with the direction from bulk Si to surface. Under illumination, electron-hole pairs generated in depletion layer of Si will be separated by the built-in electric field. As a result, the electrons sweep into the bulk Si, and the holes accumulate at the surface, which give rise to a lateral potential gradient between the illuminated and non-illuminated zones.[1-10] The photo-induced holes will diffuse, and the position of light could be determined by the difference in the photovoltage (or numbers of carriers) between the two electrodes. However, the defects, impurities and the surface states in Si would restrict the diffusion of holes, leading to a very short diffusion length. On the other hand, by integrating graphene with Si, a p-n junction (p-type of graphene was proved by the transfer curve in Figure S2) is formed with a built-in electric field from bulk Si to graphene at the interface of graphene/Si. Figure 1c show the schematic structure and energy band diagram[15,18,19] of graphene-Si heterojunction. Under illumination, the electron-holes generated in the depletion layer (the width is very high for lightly doped Si) of Si will be separated by the electric field. As a result, the holes enter the graphene and electrons enter bulk Si.[15] The separated carriers will weaken the built-in electric field across the graphene/Si junction, while the built-in electric field at the non-illuminated area will not change. This will cause a lateral potential gradient (electric field) between the illuminated and non-illuminated areas and the field drive the holes towards the electrodes (Figure 1c), which is the so-called lateral photoeffect.[1-10] Due to the high mobility[13,14] of graphene, the diffusion length of holes in graphene is very long and the efficiency of signal collection at the electrodes could be greatly improved. In addition, graphene can also absorb light and produce photo-generated carriers, with electrons entering Si and holes remain in graphene, which will extend the operating wavelength of the device due to the broadband absorption.[20,21]

A PSD with operating area of 8 mm × 8 mm was prepared in this work and 100 nm Aluminum was deposited on the back of Si acting as common-grounded electrode. Figure 1d shows the I-V characteristics of the graphene-Si junction working in the photodiode mode.[15] The graphene-Si diode has good current rectification with applied bias in dark, suggests the presence of a build-in electric field. The negative short circuit current under illumination means that the current flowing from Si to graphene, which is consistent with the hole injection into graphene.[15,18] Meanwhile, the number of holes entering graphene increases with the increase of light intensity. The carriers accumulated at Si surface (Figure 1b) or entering graphene (Figure 1c) will diffuse laterally and the numbers of carriers reach the electrodes $V_{X1}$ and $V_{X2}$ are different, which are determined by the distance between the light spot and electrodes. The photovoltage difference, $V_{X2}-V_{X1}$, can then be used to detect the position of light. Figure 1e displays the position dependence of the photovoltage difference between the two electrodes with or without graphene on Si surface. Here, "0 mm" represents the center of the device, while "-4 and 4 mm" represent the position of two electrodes. The output photovoltage difference $V_{X2}-V_{X1}$ on pure Si (inset of Figure 1e) drops rapidly to zero at ~200 μm away from electrode, suggesting that the diffusion length of carriers at Si surface is very small. On the other hand, the carriers can diffuse very long in graphene (Figure 1c) due to its high mobility and lack of surface defect states. The photovoltage of $V_{X1}$ or $V_{X2}$ drops by only ~20% in the distance of 8 mm, as shown in Figure S3a. When the position of light is close to the center of the device, the voltage difference $V_{X2}-V_{X1}$ is close to zero due to the isotropic diffusion of carriers, whereas the difference is great when the light is close to one of the electrodes. Furthermore, the almost linear dependence between the voltage difference and light position ensures that the device is competent in precisely identifying the light position. According to the diffusion length of carriers in graphene (Figure S3b), the operating area of the device could be more than 30 mm × 30 mm.

## 2.2. Position sensitive characteristics of the PSD

**Figure 2**a shows the photo-switching characteristics of the device with laser (532 nm) focused on graphene at different positions under zero bias. To avoid the error caused by the damage of graphene (winkles, broken holes produced during the transfer procedure), the ratio between the difference and the sum of the output photovoltages of the two electrodes ($V_{X2}$-$V_{X1}$)/($V_{X2}$+$V_{X1}$) was employed to display the position sensitive characteristics, which can effectively improve the linearity of the measurement (Figure S4). Figure 2b shows the position dependence of the photovoltage ratios at X-direction with different laser spot size from 5 μm to 800 μm. The exactly same linear characteristics suggest that the laser spot size does not affect the position sensitivity of our device, which is in good accordance with the characteristic of "independent of the incident light shape" of PSD. Indeed, the output signal of PSD is only determined by the gravity center position of light. The similar linear dependence of signals in both directions (X-direction and Y-direction) implies that the diffusion of holes in graphene is isotropic, which is promising for imaging or other practical applications. The non-linearity ($\delta$) is an important parameter of PSDs, which characterizes the position detection error and is usually expressed as:[8,9]

$$\delta = \frac{2 \times \sqrt{\left[\sum_{i=1}^{N}(X_i - X_i^T)^2\right]/N}}{L} \times 100\%$$

where $X_i$ is the measured position, $X_i^T$ is the actual position, $N$ is number of measured position, $L$ is the distance between the two electrodes. An acceptable device has non-linearity of less than 15%.[6] In our PSD, non-linearity of ~3% is obtained under incident power of ~10 μW. Figure 2c shows the position sensitive characteristics at different light power. This demonstrates the characteristic of "independent of the incident light power" of our device, which is another characteristic of PSD. The non-linearity of the device increases gradually with decreasing light power, due to the decrease of photovoltage for weak signals. The spatial resolution of the PSDs can also be deduced and shown in Figure S5, which shows a resolution of ~2 μm and ~0.34 μm for ~10 μW and ~100 μW light power, respectively. This is better than the value of commercial product (~6.8 μm and ~0.68 μm for ~10 μW and ~100 μW light). Although the output signal of each electrode ($V_{X1}$ and $V_{X2}$) is related to the light position, the sum of photovoltages ($V_{X1}+V_{X2}$) from the two electrodes keeps almost constant (Figure S6). Figure 2d displays the sum of photovoltages ($V_{X1}+V_{X2}$) with the increase of light power, which shows almost linear dependence. This demonstrates that our PSD could also be used for the detection of incident light power, in addition to the positions.

**2.3. High performance of the PSD**

In order to explore the capability of our device to detect weak light signals, position sensitive characteristics of the PSD at different light powers are carried out. **Figure 3**a displays the power dependence of the photovoltage difference. Evidently, the values increase linearly with power from a dozen of nW to almost 100 μW. The linear region is very broad compared to the operating power range of Si based PSD from several μW to a few hundred of μW. The phenomenon of saturation at high power can be interpreted by the balance between the internal built-in electric field and the accumulation of photoexcited carriers. The power detection limit of our PSD is ~17 nW, with its linear position sensitive characteristic shown in Figure 3b. This demonstrates that our PSD can be used for weak signal detection down to nWs level, which is very promising in applications. Figure 3c shows the transient response of the device by using an acoustic optical modulator with frequency of 10 kHz to switch the light

(633nm, ~40 μW). The rise ($\tau_{on}$) and fall ($\tau_{off}$) time (light located at 10 μm away from the electrode $V_{X1}$) are ~430 and ~450 ns, respectively, where the curves are fitted using a single exponential function. The dependence of rise time as a function of light position is shown in Figure 3d. The increases of response time with distance could be understood as the increase of transit time of holes in graphene with increasing distance between light and the electrode. Despite the increase, the response time of a few μs is still fast enough for various applications. The fast response is attributed to fast separation of photoexcited carriers at graphene-Si junctions and the high mobility of graphene for carrier transportation.

**2.4. Characteristics of the PSD for infrared light**

Infrared PSDs are especially important for military applications, for example the laser guiding systems. However, the photosensitive material of the conventional Si-based PSD is only Si, which limits the operating wavelength in the range of 300-1100 nm. The introduction of graphene in our PSD will extend the operating wavelength due to the broadband absorption caused by the special zero bandgap structure of graphene.[20,21] **Figure 4**a depicts the photoresponse of the device to 1319 nm laser. Stable and fast response implies the capability of the device to near infrared. Figure 4b demonstrates the position sensitive characteristics of the PSD to infrared light under different powers, which are linear and power independent. However, the minimum operating power of 1319 nm infrared is ~10 μW, due to the weak light absorption of graphene.[20] The output difference, $V_{X2}$-$V_{X1}$, of the PSD as a function of power is shown in Figure 4c. It shows good linearity and suggests the broad operating power range of our PSD for near infrared light. The position sensitive characteristic of the PSD to 1550 nm infrared (4 mW) was also carried out. The similar ultrafast and stable response and the linear dependence are shown in Figure 4d, but with a minimum operating power of mW level. The weak photo response to 1550 nm light might be related to the energy barrier between graphene and Si.

**2.5. Application of the PSD**

To prove the capability of position detection of graphene-Si based PSD, an 8 mm × 8 mm device was prepared and encapsulated, as shown in **Figure 5**a. When the laser beam (633 nm, ~40 μW) moves along the trajectory of a square shape within the operating area (Figure 5a), the real-time position of the light spot can be obtained through the output of the two pairs (X and Y) of electrodes. Figure 5b exhibits the experimentally extracted trajectory of the light spot, which agrees well with the programmed pattern (white dashed square). The experimental errors of the two dimensional PSD are larger than that of one dimensional measurement (Figure 2b), which is probably due to the influence of the electrodes and lack of system calibration. The adoption of pillow-shaped electrodes can effectively reduce the error for achieving high-precision detection. It should be emphasized that our PSD is a passive device, which means there is no power consumption during the measurement, while conventional Si P-i-N based PSDs would require a bias voltage. This advantage could be promising for portable and integrated devices.

**3. Conclusion**

In conclusion, we present a high-performance passive PSD based on the graphene-Si hybrid structure. The PSD characterizes excellent position sensitivity to weak light at nWs level. More importantly, it shows very fast response speed and low degree of non-linearity of ~3%, and extends the operating wavelength to near IR. The characteristic of our PSD is also independent on the size and power of the light spot, and be used for detecting the power of incident light besides its position. This work therefore provides a new opportunity for PSDs with ultrahigh sensitivity and broadband response.

## 4. Experimental Section

*Fabrication of graphene-Si based PSDs*: Monolayer graphene film grown on copper foil catalyst surface was transferred onto lightly n-doped Si ($\rho$ = 1-10 $\Omega$ cm) substrate.[22] Two pairs of Ni (5 nm)/Au (50 nm) electrodes ($V_{X1}$, $V_{X2}$ and $V_{Y1}$, $V_{Y2}$) were deposited on graphene for signals collection by using mask and thermal evaporation (TPRE-Z20-IV). In addition, 100 nm Al was deposited on the bottom of Si substrate acting as common-grounded electrode.

*Device characterization*: The electrical characteristics were measured using a Keithley 2612 analyzer. The photovoltage were measured using focused laser beams (spot size ~1 µm) with wavelength of ~532, ~633, ~1319 and ~1550 nm. In the spot size dependent experiment, various spot sizes were employed by controlling the focus of the objective lens. The position sensitive characteristics were measured by moving the device with a two dimensional motorized stage. In the response time measurement, light (~633 nm) was modulated with an acoustic optical modulator (R21080-1DS) at frequency of 10 kHz. A digital storage oscilloscope (Tektronix TDS 1012, 100 MHz/1GS/s) is used to measure the transient response of photocurrent. In the demonstration of two dimensional PSD, the two-dimensional motorized stage and two Keithley 2612 analyzers were controlled by the labview software simultaneously to obtain the real-time position of the laser. All the measurements were performed in air at room temperature.

References

[1]  S. Arimoto, H. Yamamoto, H. Ohno, H. Hasegawa, *J. Appl. Phys.* **1984**, *57*, 4778.

[2]  G. Lucovsky, *J. Appl. Phys.* **1960**, *31*, 1088.

[3]  E. Fortunato, G. Lavareda, M. Vieira, R. Martins, *Rev. Sci. Instrum.* **1994**, *65*, 3784.

[4]  R. Martins, E. Fortunato, *Rev. Sci. Instrum.* **1995**, *66*, 2927.

[5]  M. Vieira, *Appl. Phys. Lett.* **1997**, *70*, 220.

[6]  E. Fortunato, G. Lavareda, R. Martins, F. Soares, L. Fernandes, *Sens. Actuators A* **1996**, *51*, 135.

[7]  B. F. Levine, R. H. Willens, C. G. Bethea, D. Brasen, *Appl. Phys. Lett.* **1986**, *49*, 1537.

[8]  S. Q. Xiao, H. Wang, Z. C. Zhao, Y. Z. Gu, Y. X. Xia, Z. H. Wang, *Opt. Express* **2008**, *16*, 3798.

[9]  J. Henry, J. Livingstone, *Phys. Status Solidi A* **2011**, *208*, 1718.

[10] C. Q. Yu, H. Wang, Y. X. Xia, *Appl. Phys. Lett.* **2009**, *95*, 263506.

[11] P. F. Fonteint, P. Hendrikst, J. H. Woltert, A. Kucernakg, R. Peat, D. E. Williams, *Sci. Technol.* **1989**, *4*, 837.

[12] C. G. Mattsson, G. Thungstrom, H. Rodjegard, K. Bertilsson, H. E. Nilsson, H. Martin, *IEEE Sens. J.* **2009**, *9*, 1633.

[13] K. S. Novoselov, A. K. Geim, S. V. Morozov, D. Jiang, Y. Zhang, S. V. Dubonos, I. V. Grigorieva, A. A. Firsov, *Science* **2004**, *306*, 666.

[14] K. S. Novoselov, A. K. Geim, S. V. Morozov, D. Jiang, M. I. Katsnelson, I. V. Grigorieva, S. V. Dubonos, A. A. Firsov, *Nature* **2005**, *438*, 197.

[15] X. An, F. Liu, Y. J. Jung, S. Kar, *Nano Lett.* **2013**, *13*, 909.

[16] Z. H. Ni, L. A. Ponomarenko, R. R. Nair, R. Yang, S. Anissimova, I. V. Grigorieva, F. Schedin, P. Blake, Z. X. Shen, E. H. Hill, K. S. Novoselov, A. K. Geim, *Nano Lett.* **2010**, *10*, 3868.

[17] M. Akatsuka, K. Sueoka, *Jpn. J. Appl. Phys.* **2001**, *40*, 1240.

[18] X. Li, M. Zhu, M. Du, Z. Lv, L. Zhang, Y. Li, Y. Yang, T. Yang, X. Li, K. Wang, H. Zhu, Y. Fang, *Small* **2016**, *12*, 595.

[19] X. Li, H. Zhu, K. Wang, A. Cao, J. Wei, C. Li, Y. Jia, Z. Li, X. Li, D. Wu, *Adv Mater* **2010**, *22*, 2743.

[20] R. R. Nair, P. Blake, A. N. Grigorenko, K. S. Novoselov, T. J. Booth, T. Stauber, N. M. R. Peres, A. K. Geim, *Science* **2008**, *320*, 1308.

[21] K. F. Mak, L. Ju, F. Wang, T. F. Heinz, *Solid State Commun.* **2012**, *152*, 1341.

[22] X. Li, W. Cai, J. An, S. Kim, J. Nah, D. Yang, R. Piner, A. Velamakanni, I. Jung, E. Tutuc, *Science* **2009**, *324*, 1312.

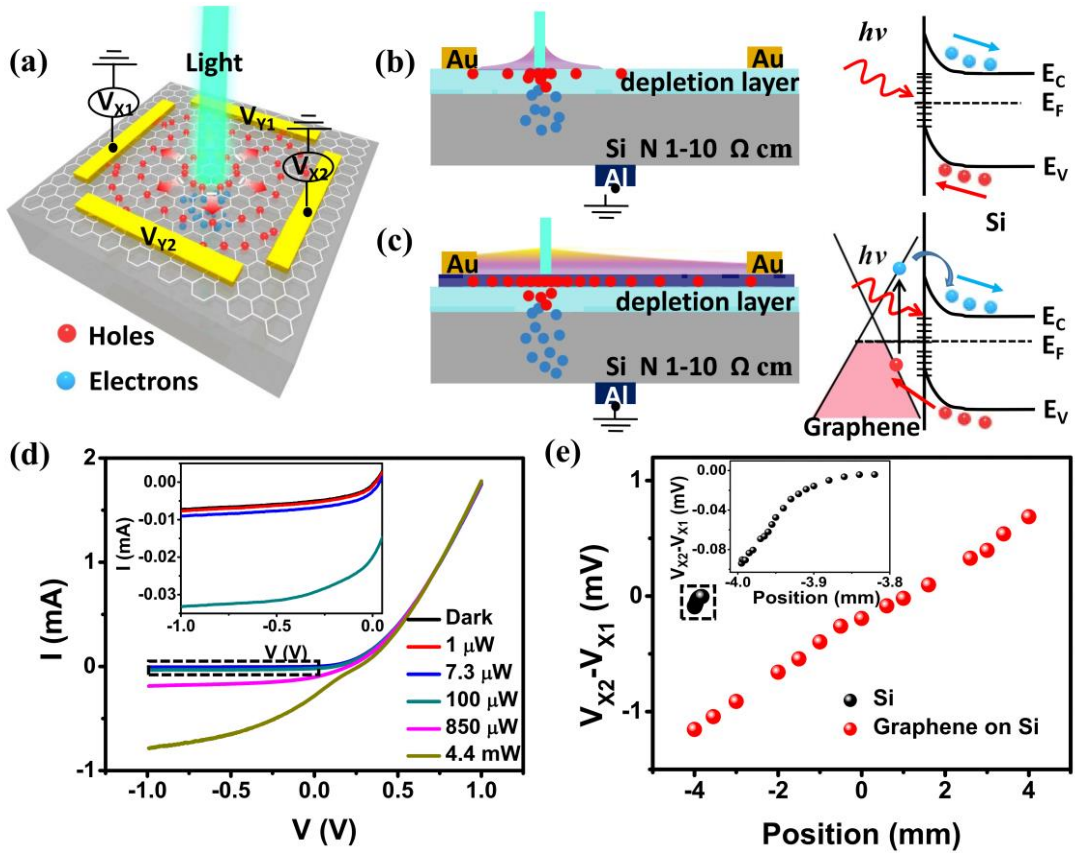

**Figure 1** Schematic of PSD based on graphene-Si heterojunction. a) Schematic diagram of the graphene device on lightly n-doped Si substrate. The photo-generated holes sweep into graphene under the built-in electric field, and then diffuse laterally, while the electrons enter the bulk Si. b,c) Schematic structure and energy band diagrams of devices without (b) and with (c) graphene on Si substrate. The pinning effect causes the surface band to bend upwards, leading to a built-in electric field from bulk Si to the surface. The diffusion length of holes in graphene is much longer than that in Si. d) Electrical characteristics of the graphene/Si junction as a photodiode in dark and under illumination of different powers. e) The spatial dependence of the difference of output photovoltage ($V_{X2}-V_{X1}$) measured from pure Si and graphene/Si, as shown in (b) and (c), respectively. Set the middle of the two electrodes ($V_{X2}, V_{X1}$) as the coordinate origin, and the interface of graphene-electrode $V_{X1}$ and $V_{X2}$ as -4 and 4 mm, respectively. (532 nm laser with spot size ~1 µm)

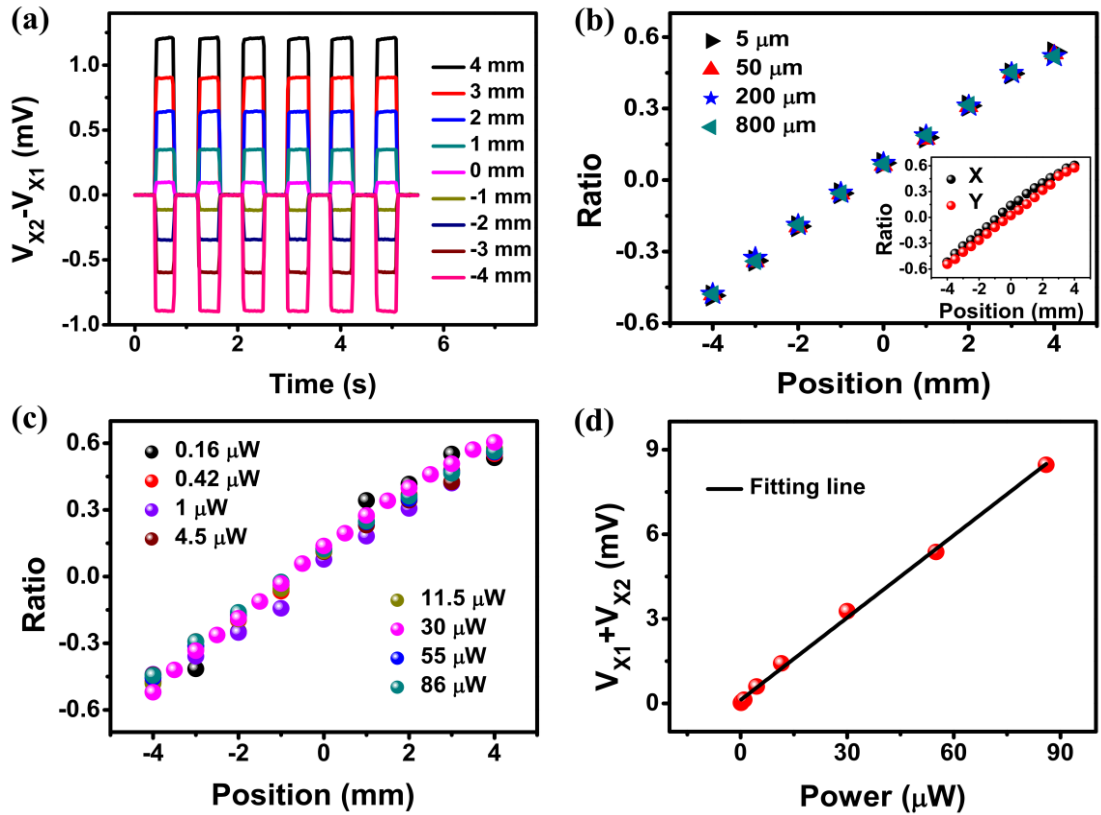

**Figure 2** Position sensitive characteristics of the graphene/Si based PSD. a) Photo-switching characteristics of the PSD with 820 nW incident light (532 nm) focusing at different positions. b) The ratio $[(V_{X2}-V_{X1})/(V_{X1}+V_{X2})]$ as a function of the laser position with different spot sizes. The inset plots the position sensitive characteristics of both directions (X and Y). c) The position sensitive characteristics under different incident light power. d) The power dependence of the sum of the output voltages $(V_{X1}+V_{X2})$ in X direction.

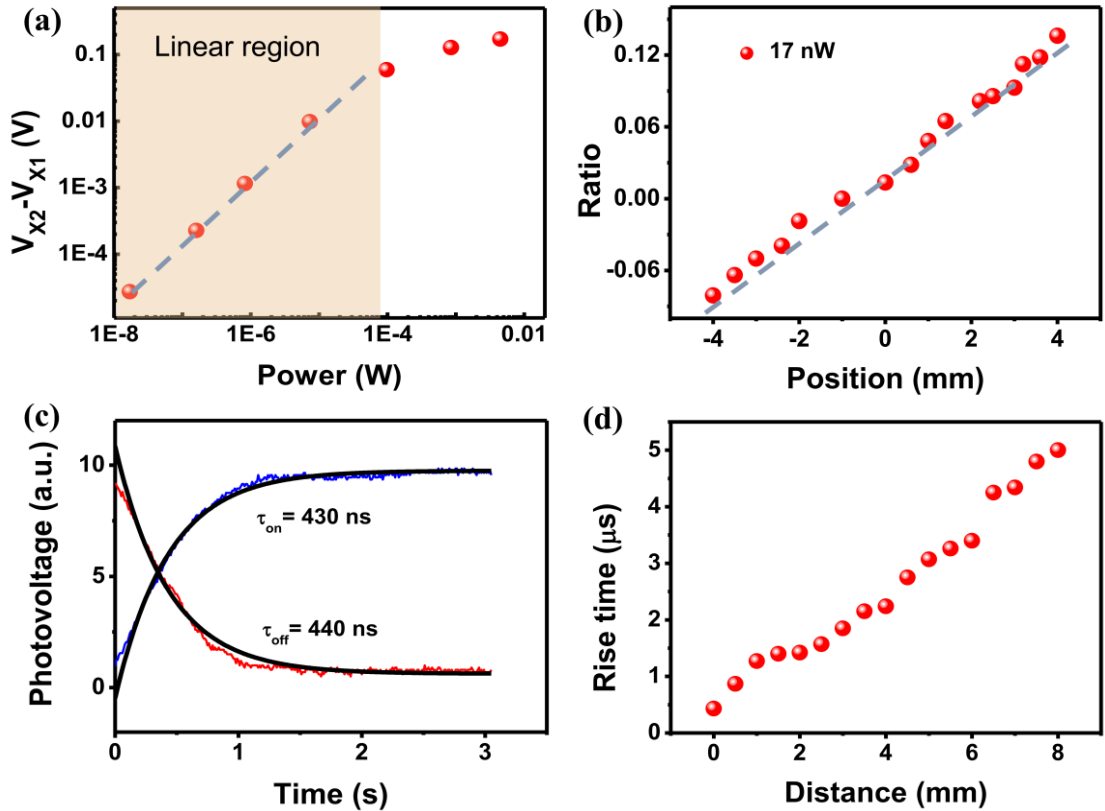

**Figure 3** The performance of graphene/Si based PSD. a) The power dependence of the difference of the output voltages ($V_{X2}-V_{X1}$) when the light spot is focused on the position of 1 μm away from the electrode $V_{X2}$. b) The position sensitive characteristics under weak incident light of ~17 nW. c) The transient response of the device when the light (633 nm laser, at the position of 1 μm away from the electrode $V_{X1}$) is switched on or off by an acoustic optical modulator with frequency of 10 kHz. d) The distance dependence of the rise time of the device.

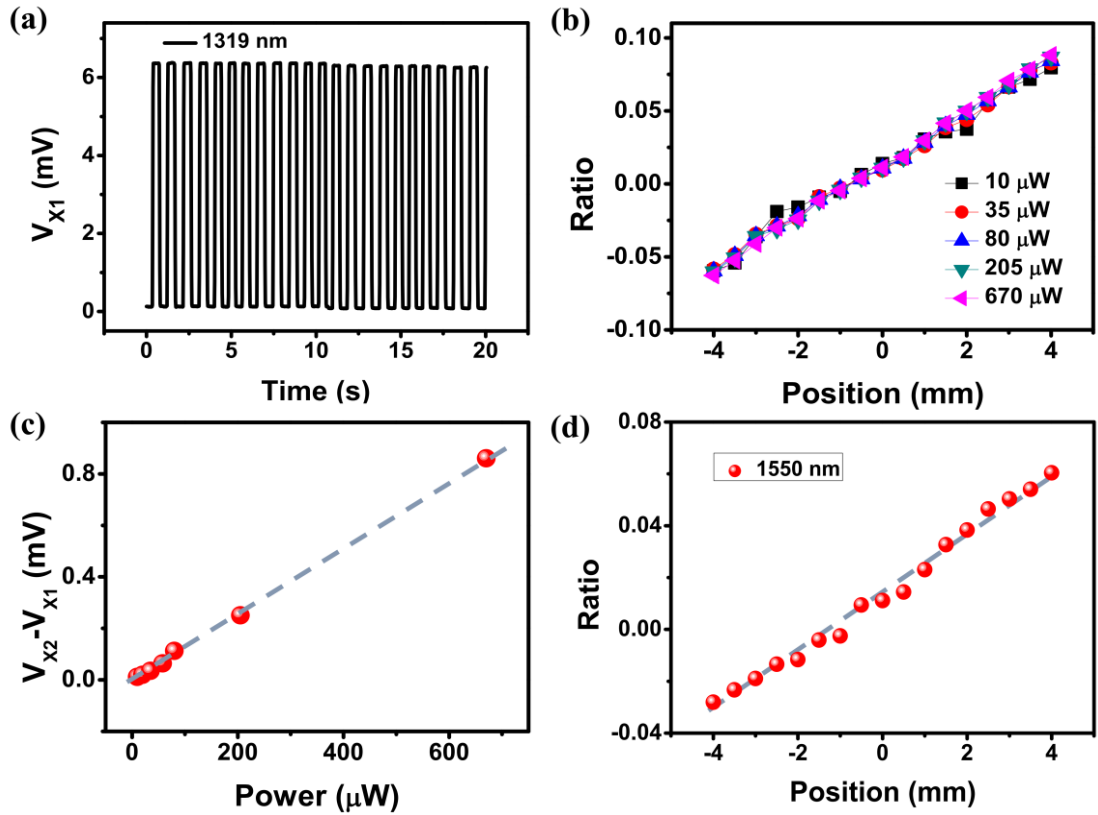

**Figure 4** The position sensitive characteristics of the PSD for infrared light. a) Time-dependent output photovoltage of the PSD with periodic chopping of infrared light source (1319 nm, 670 μW). b) The position sensitive characteristics under different power of infrared light (1319 nm). c) The difference of the output voltages ($V_{X2}-V_{X1}$) as a function of power when the light spot is focused at 1 μm away from the electrode $V_{X2}$ (1319 nm). d) The position sensitive characteristics for 1550 nm infrared light (4 mW).

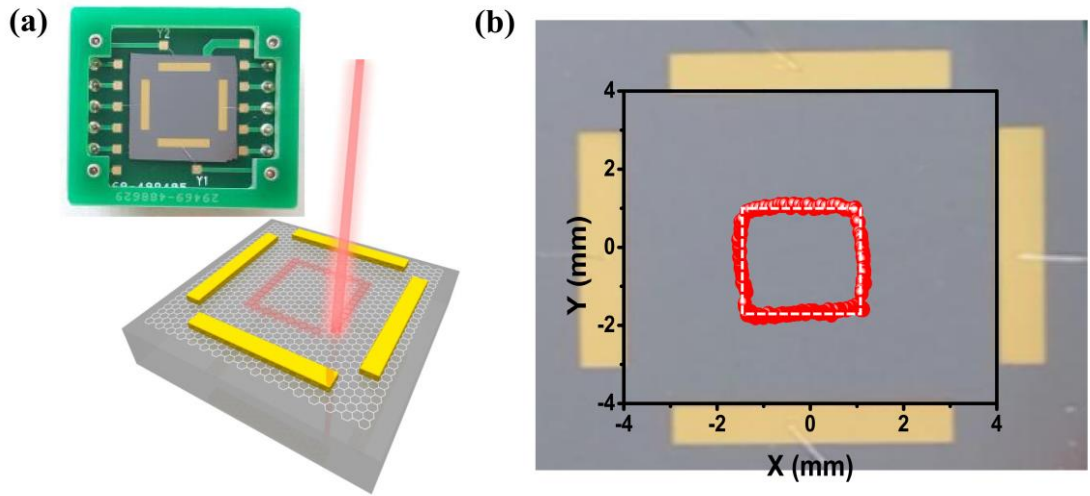

**Figure 5** The demonstration of two dimensional PSD. a) The optical image of an 8 mm × 8 mmPSD and schematic diagram of the movement of light in the operating area. b) The measured trajectory of the laser (633 nm, ~40 µW). The white dotted line represents the actual position.

# Supporting Information

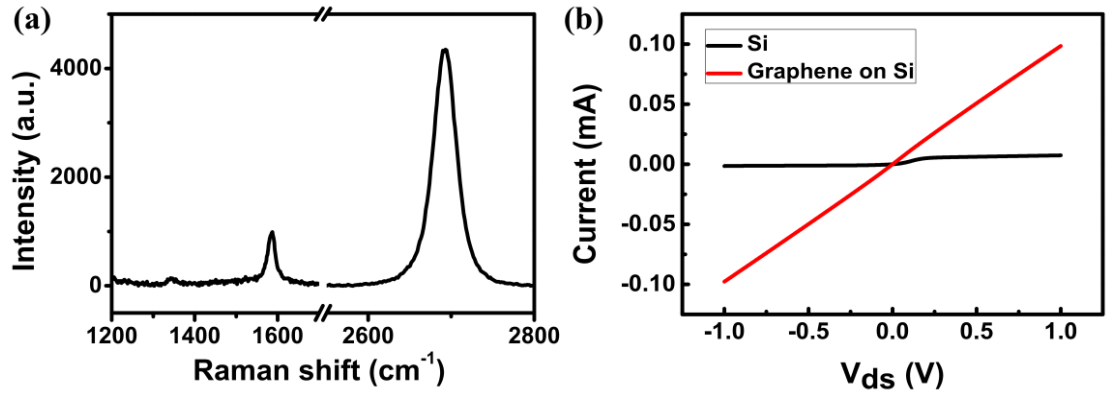

**Figure S1** a) Raman spectrum of graphene on lightly n-doped Si (1-10 Ω cm). b) I-V curves of the pure Si and graphene on lightly n-doped Si.

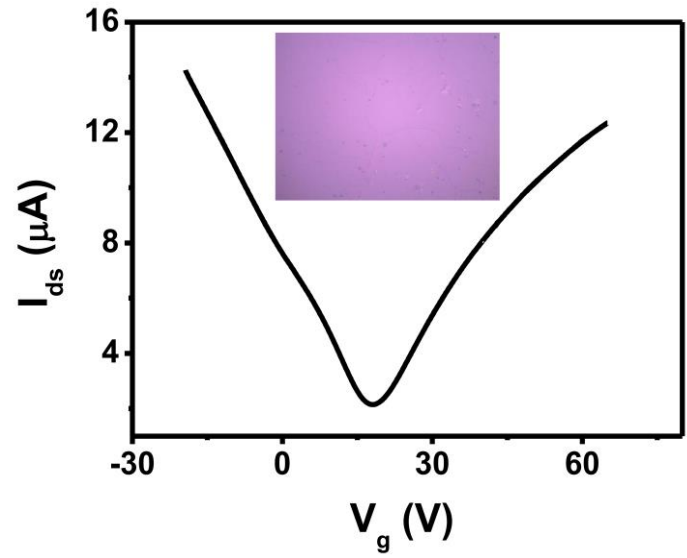

**Figure S2** The transfer curve of CVD graphene device on 300 nm $SiO_2$/Si substrate. The mobility of graphene is ~4000 $cm^2$/Vs. $V_{ds}$ = 0.01 V. The inset is part of the optical picture of the CVD graphene.

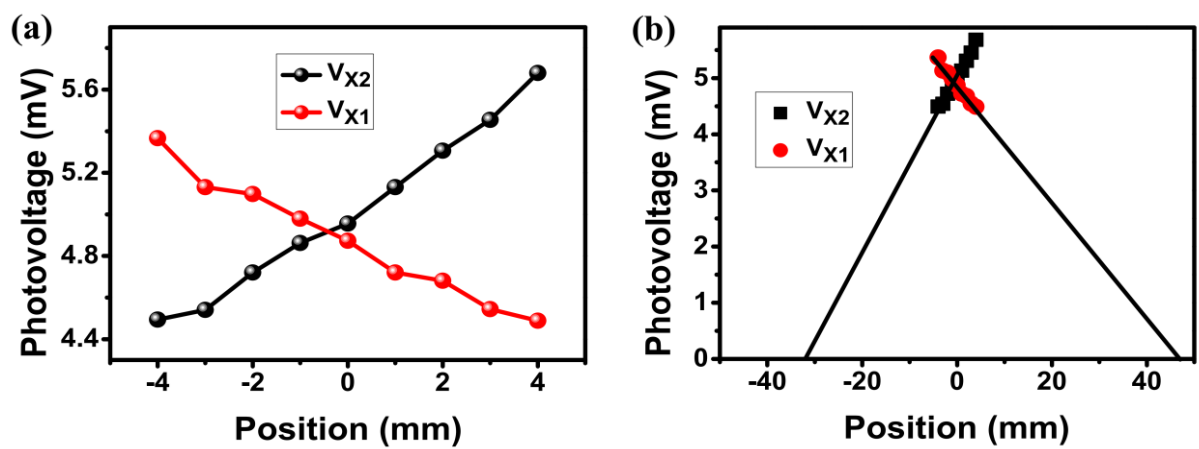

**Figure S3** a) The output photovoltage $V_{X1}$ and $V_{X2}$ as a function of position. b) Extension lines of the output photovoltage $V_{X1}$ and $V_{X2}$. According to the diffusion length of carriers in graphene, the operating area of the device could be more than 30 mm × 30 mm.

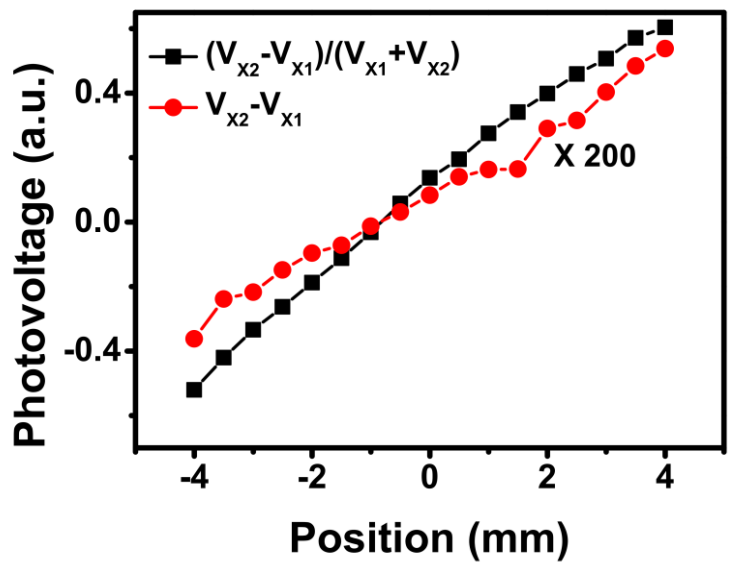

**Figure S4** The position sensitive characteristics by using the difference ($V_{X2}-V_{X1}$) and the ratio [($V_{X2}-V_{X1}$)/($V_{X1}+V_{X2}$)].

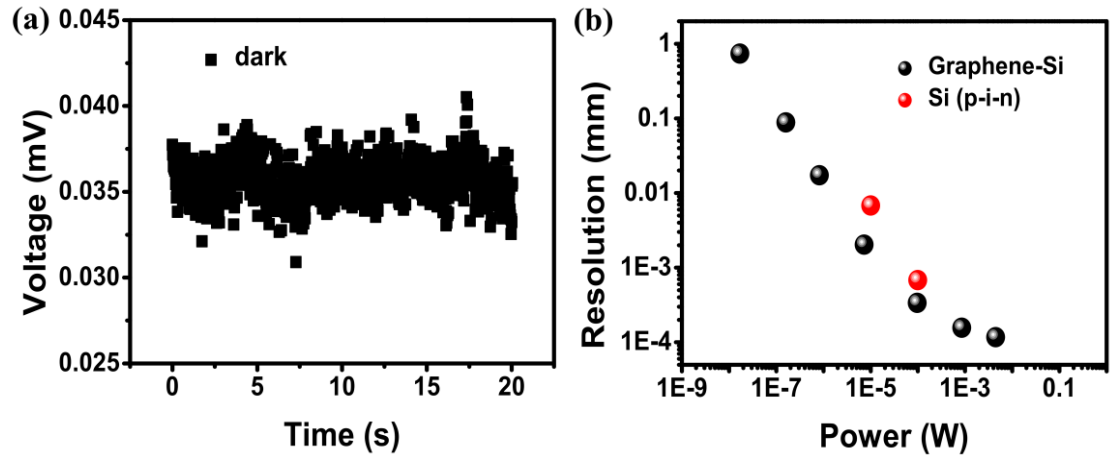

**Figure S5** a) The dark current waveform of the graphene-Si based PSD, the noise current is ~5 μV. b) Power dependence of the resolution of graphene-Si PSD and Si (p-i-n) based PSD.

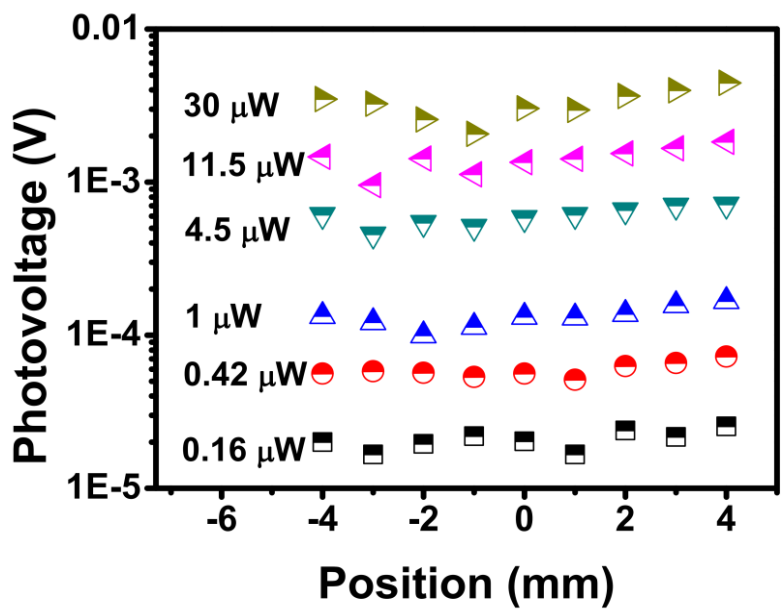

**Figure S6** The position dependence of the sum of photovoltage $V_{X1}+V_{X2}$ under different incident light power.

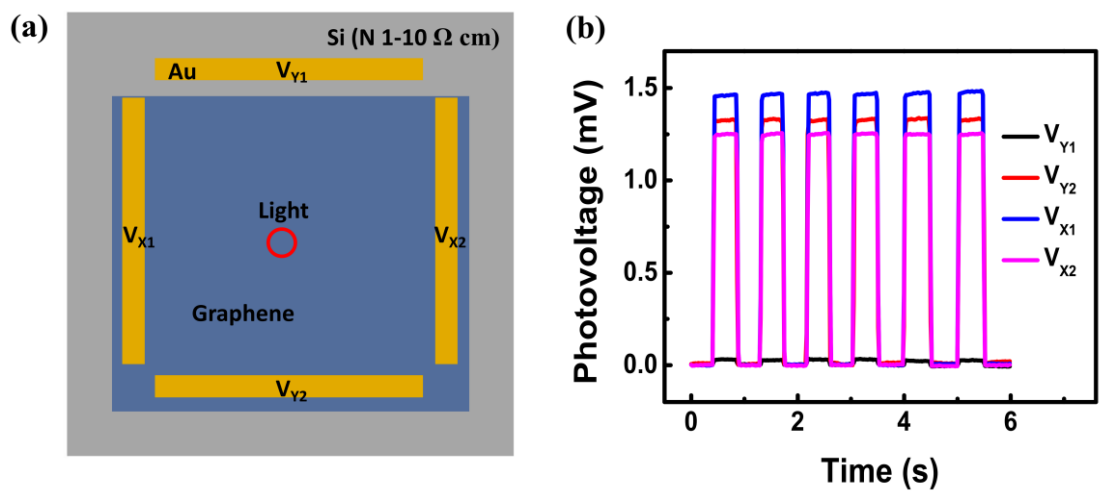

**Figure S7** a) Schematic diagram of the graphene-Si based PSD with one electrode ($V_{Y1}$) on lightly n-doped Si substrate. b) The output photovoltage of all four electrodes.

To further prove the role of graphene in the PSD, a comparative experiment was carried out. Three electrodes were placed on graphene and the other one is on Si. The output signal of each electrode was recorded in Fig. S6 when light is illuminated on graphene region. The signals of the three electrodes on graphene are significant, while that of the electrode on Si is negligible, which fully support that the carriers diffuse in graphene instead of Si.

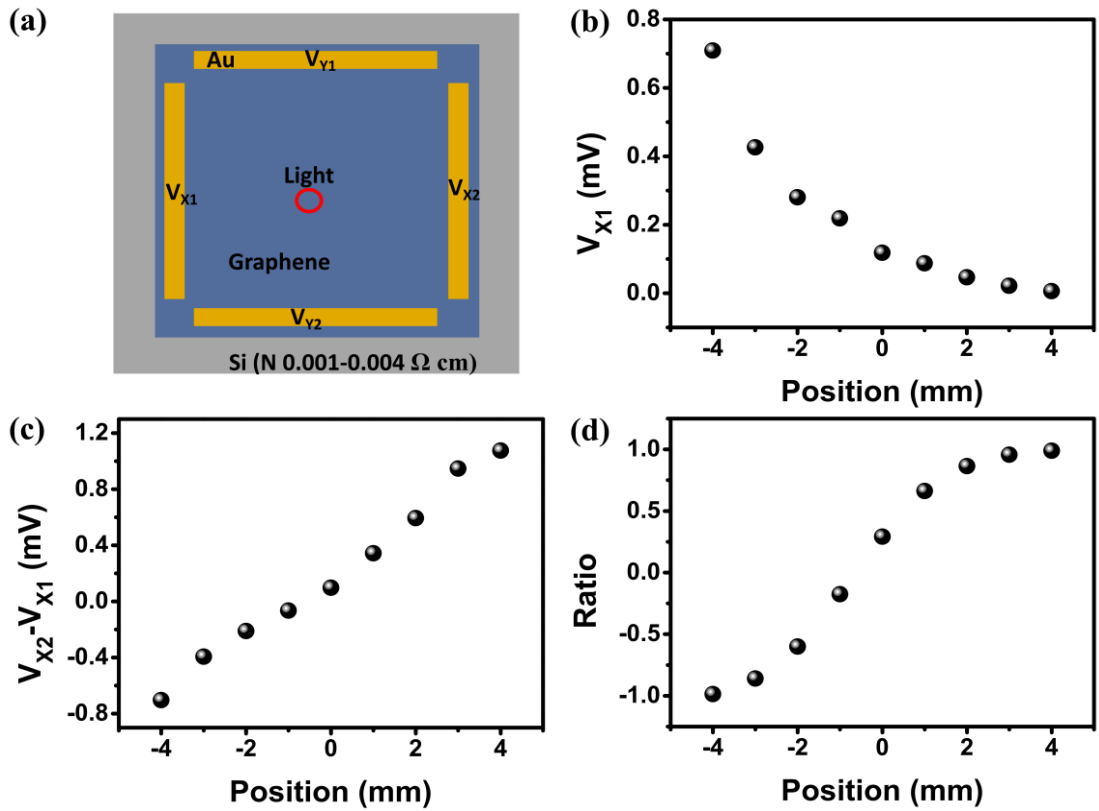

**Figure S8** a) Schematic diagram of the PSD based graphene on heavily n-doped Si substrate. b,c,d)The output photovoltage $V_{X1}$ (b), difference $V_{X2}$-$V_{X1}$ (c) and ratio $(V_{X2}-V_{X1})/(V_{X1}+V_{X2})$ (d) as a function of position.